\documentclass{emulateapj}

\shorttitle{Discovery of gamma-ray emission from PKS~1454-354}
\shortauthors{The \emph{Fermi}/LAT Collaboration}

\begin{document}

%% LaTeX will automatically break titles if they run longer than
%% one line. However, you may use \\ to force a line break if
%% you desire.

\title{\emph{Fermi}/LAT discovery of gamma-ray emission\\from the flat-spectrum radio quasar PKS~1454-354}

%% Use \author, \affil, and the \and command to format
%% author and affiliation information.
%% Note that \email has replaced the old \authoremail command
%% from AASTeX v4.0. You can use \email to mark an email address
%% anywhere in the paper, not just in the front matter.
%% As in the title, use \\ to force line breaks.

%\author{The Fermi-LAT Collaboration}
%\affil{SLAC, Stanford, California, USA}

% Author list updated on 3 March 2009: 181 LAT authors + 5 external authors.

\author{
A.~A.~Abdo\altaffilmark{1,2}, 
M.~Ackermann\altaffilmark{3}, 
W.~B.~Atwood\altaffilmark{4}, 
M.~Axelsson\altaffilmark{5,6}, 
L.~Baldini\altaffilmark{7}, 
J.~Ballet\altaffilmark{8}, 
G.~Barbiellini\altaffilmark{9,10}, 
D.~Bastieri\altaffilmark{11,12}, 
B.~M.~Baughman\altaffilmark{13}, 
K.~Bechtol\altaffilmark{3}
% ...
}
% ...
\altaffiltext{1}{National Research Council Research Associate}
\altaffiltext{2}{Space Science Division, Naval Research Laboratory, Washington, DC 20375}
\altaffiltext{3}{W. W. Hansen Experimental Physics Laboratory, Kavli Institute for Particle Astrophysics and Cosmology, Department of Physics and SLAC National Laboratory, Stanford University, Stanford, CA 94305}
\altaffiltext{4}{Santa Cruz Institute for Particle Physics, Department of Physics and Department of Astronomy and Astrophysics, University of California at Santa Cruz, Santa Cruz, CA 95064}
\altaffiltext{5}{The Oskar Klein Centre for Cosmo Particle Physics, AlbaNova, SE-106 91 Stockholm, Sweden}
\altaffiltext{6}{Department of Astronomy, Stockholm University, SE-106 91 Stockholm, Sweden}
\altaffiltext{7}{Istituto Nazionale di Fisica Nucleare, Sezione di Pisa, I-56127 Pisa, Italy}
\altaffiltext{8}{Laboratoire AIM, CEA-IRFU/CNRS/Universit\'e Paris Diderot, Service d'Astrophysique, CEA Saclay, 91191 Gif sur Yvette, France}
\altaffiltext{9}{Istituto Nazionale di Fisica Nucleare, Sezione di Trieste, I-34127 Trieste, Italy}
\altaffiltext{10}{Dipartimento di Fisica, Universit\`a di Trieste, I-34127 Trieste, Italy}
\altaffiltext{11}{Istituto Nazionale di Fisica Nucleare, Sezione di Padova, I-35131 Padova, Italy}
\altaffiltext{12}{Dipartimento di Fisica ``G. Galilei", Universit\`a di Padova, I-35131 Padova, Italy}
\altaffiltext{13}{Department of Physics, Center for Cosmology and Astro-Particle Physics, The Ohio State University, Columbus, OH 43210}
% ...

\begin{abstract}
We report the discovery by the Large Area Telescope (LAT) onboard the \emph{Fermi Gamma-ray Space Telescope} of high-energy $\gamma-$ray (GeV) emission from the flat-spectrum radio quasar PKS~1454-354 ($z=1.424$). On $4$~September~$2008$ the source rose to a peak flux of $(3.5\pm 0.7)\times 10^{-6}$~ph~cm$^{-2}$~s$^{-1}$ ($E>100$~MeV) on a time scale of hours and then slowly dropped over the following two days. No significant spectral changes occurred during the flare. \emph{Fermi}/LAT observations also showed that PKS~1454-354 is the most probable counterpart 
of the unidentified EGRET source 3EG~J$1500-3509$. Multiwavelength measurements performed during the following days ($7$~September with \emph{Swift}; $6-7$~September with the ground-based optical telescope \emph{ATOM}; $13$~September with the \emph{Australia Telescope Compact Array}) resulted in radio, optical, UV and X-ray fluxes greater than archival data, confirming the activity of PKS~1454-354. 
\end{abstract}


\section{Introduction}
PKS~1454-354 was discovered in the Molonglo Radio Telescope Survey at $408$~MHz (Large et al. 1981) and later identified as a flat-spectrum radio quasar at $z=1.424$ (Jackson et al. 2002, Hook et al. 2003). VLBA observations at $2.3$ and $8.4$ GHz showed a compact core with a weak one-sided jet (Petrov et al. 2005). 

Interest in this source was generated by the detection by EGRET onboard the \emph{Compton Gamma-Ray Observatory} (CGRO) of the unidentified source 3EG~J1500-3509 (Hartman et al. 1999). Mattox et al. (2001) tentatively indicated PKS~1454-354 as the only possibile association with the EGRET unidentified source, but the association probability was very low ($0.027$). However, they noted that the EGRET probability contours were not closed on the eastern side and they suggested that this could be due to a contaminating source. The same possibility was proposed by Tornikoski et al. (2002) on the basis of millimeter observations, while Sowards-Emmerd et al.~(2004) suggested that PKS~1454-354 could be one of the two likely counterparts of the unidentified EGRET source (the other is PMN~J1505-3432). 

The answer to this conundrum comes from the \emph{Fermi} satellite and is presented here. The wide field of view ($\sim 2.4$~sr) of the Large Area Telescope (LAT, Atwood et al. 2008) coupled with its high sensitivity (more than an order of magnitude better than EGRET) and spatial resolution (of the order of a few tens of arcminutes for $E>1$~GeV), make it possible to monitor the whole sky every two orbits ($\sim 3$~hours). During these scans, on $4$~September~$2008$, a strong $\gamma-$ray flare was observed from the direction of PKS~1454-354 (Marelli 2008). Such alert triggered follow-up observations with \emph{Swift} (Gehrels et al. 2004), performed on $7$~September, with the \emph{Australia Telescope Compact Array} (\emph{ATCA}), done on $13$~September and with the \emph{Automated Telescope for Optical Monitoring} (\emph{ATOM}), which started the observations on September~$6$.

Here we report on the analysis of \emph{Fermi}/LAT, \emph{Swift}, \emph{ATCA} and \emph{ATOM} data collected during the period of $\gamma-$ray activity and follow-up observations (Sect. 2), together with a comparison with archival data and a preliminary interpretation of the results (Sect. 3).

\section{Data Analysis}
\subsection{Fermi/LAT}
The Large Area Telescope (LAT) onboard the {\em Fermi} satellite detects photons with energy from $20$~MeV to $>300$~GeV through the conversion to electron-positron pairs in a silicon tracker (Atwood et al. 2008).  

The data have been analyzed using {\tt Science Tools v 9.7}, a software package dedicated to the \emph{Fermi}/LAT data analysis\footnote{http://fermi.gsfc.nasa.gov/ssc/data/analysis/software/}.  Events that have the highest probability of being photons, labeled ``diffuse'' in the \texttt{Science Tools}, coming from zenith angles $< 105^\circ$ (to avoid Earth's albedo) were selected. The diffuse emission from the Milky Way has been subtracted by using specific maps based on the GALPROP model (Strong et al. 2004a,b), while the extragalactic diffuse emission and residual instrumental backgrounds were modeled together as an isotropic component with power-law spectral shape. 

Photons were extracted from a region with a $10^{\circ}$ radius centered on the coordinates of the radio position of PKS~1454-354 ($\alpha=224^{\circ}.36$ and $\delta=-35^{\circ}.65$, J$2000$) and analyzed with an unbinned likelihood (Cash 1979, Mattox et al. 1996), which is implemented in the \texttt{LAT Science Tools} as the \texttt{gtlike} task. Because of calibration uncertainties at low energies, data were selected with energies above $200$~MeV, with the highest-energy $\gamma-$ray being recorded at $\sim 5$~GeV.

\begin{figure}
\centering
\includegraphics[scale=0.4]{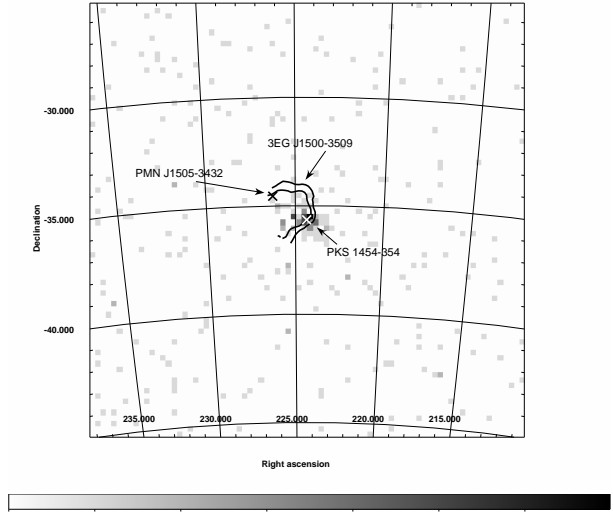}
\caption{\emph{Fermi}/LAT counts map ($0.2-300$~GeV) of the region centered on PKS~1454-354 (indicated with a white cross) as observed on September $4$, $2008$. The $95$\% and $99$\% probability contours of 3EG~J1500-3509 are superimposed. PMN~J1505-3432, the other possible counterpart suggested by Sowards-Emmerd et al. (2004), is indicated with a black cross. Epoch of coordinates is J2000.}
\label{onedaymap}
\end{figure}

On $4$~September~$2008$ (the day of the discovery, Marelli 2008), with the inclusion of all the photons accumulated over the entire day (net on source time $38$~ks), the best fit position is $\alpha=224^{\circ}.55$ and $\delta=-35^{\circ}.23$ (J$2000$), with a 95\% error radius of $0^{\circ}.48$ and detection test statistic $TS = 445$ (a $21\sigma$ detection, see Mattox et al. 1996 for the definition of $TS$). Figure~\ref{onedaymap} shows the LAT counts map with the peak of counts consistent with the radio position of PKS~1454-354. The $95$ and $99\%$ contours of source location probability for 3EG~J1500-3509 from the Third EGRET Catalog (Hartman et al. 1999) are superimposed, indicating that indeed PKS~1454-354 is one contributor to the EGRET source. No indication of another source inside the EGRET probability contours is present in the counts map obtained by integrating the LAT data over the period from $1$~August~$2008$~00:00~UTC to $1$~September~$2008$~00:00~UTC (net on source time $1.1$~Ms; hereafter indicated as the ``1-month data''), with an upper limit of $2\times 10^{-8}$~ph~cm$^{-2}$~s$^{-1}$ ($90\%$ confidence level, $0.2-300$~GeV energy band) for PMN~J1505-3432, the other candidate counterpart according to Sowards-Emmerd et al. (2004). PKS~1454-354 is detected in the 1-month data with a flux $(4.5\pm 0.9)\times 10^{-8}$~ph~cm$^{-2}$~s$^{-1}$, consistent with the flux of 3EG~J1500-3509 reported in the Third EGRET Catalog (Hartman et al. 1999), suggesting that the FSRQ can be the likely counterpart and the distortion of the EGRET probability contours is due to some residual background or to some residual gamma-ray emission from the source PMN~J1505-3432 located at the boundaries of the EGRET source.

\begin{figure}
\centering
\includegraphics[angle=270,scale=0.35]{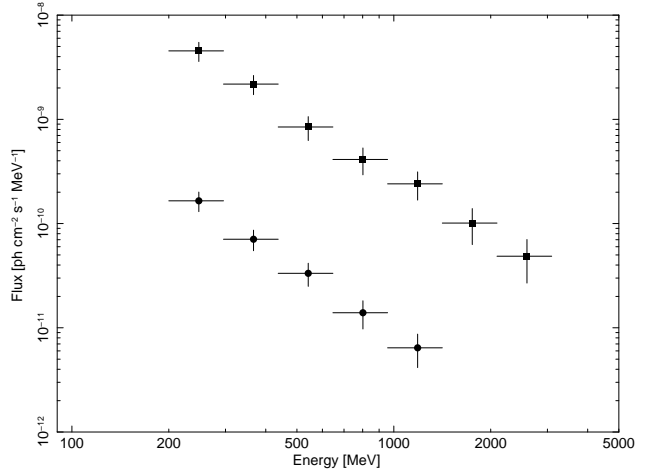}
\caption{\emph{Fermi}/LAT unfolded spectra of PKS~1454-354 from data collected on $4$~September~$2008$ (source ontime $38$~ks; filled squares) and in August $2008$ (1-month data set, source ontime $1.1$~Ms, filled circles). See the text for details on systematic errors.}
\label{spectra}
\end{figure}

\begin{figure}
\centering
\includegraphics[angle=270,scale=0.35]{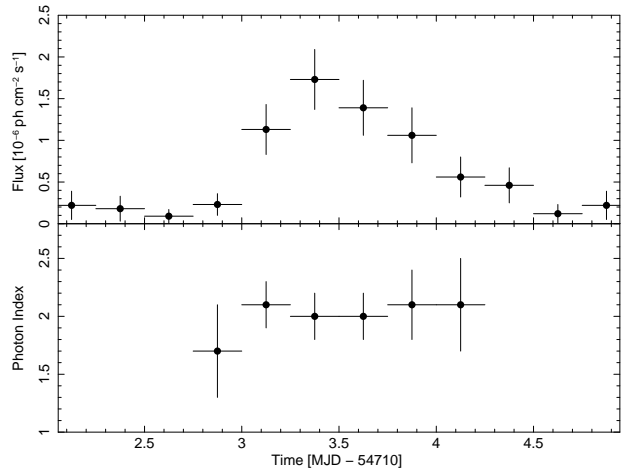}
\caption{\emph{Fermi}/LAT lightcurve in the $0.2-300$~GeV energy band with $6$-hours time bins (\emph{upper panel}) and photon index (\emph{lower panel}). Photon indices obtained from the fit with $TS < 20$ are not reported. See the text for details on systematic errors.}
\label{lcurves}
\end{figure}

\begin{table*}
\begin{center}
\scriptsize
\caption{Summary of the spectral fitting of the \emph{Fermi}/LAT data. See the text for details on systematic errors.\label{tab:summary}}
\begin{tabular}{lcrccl}
\tableline\tableline
Time(\tablenotemark{a}) & $F_{0.2-300\rm GeV}$(\tablenotemark{b}) & Fit(\tablenotemark{c}) & $F_{0.1-300\rm GeV}$(\tablenotemark{d}) & TS & Notes\\
\tableline
$4$~Sept.~00-24 & $1.29\pm 0.15$ & PL: $\Gamma=2.0\pm 0.1$ & $2.58\pm 0.30$ & $437$ & Day of the discovery \\
$4$~Sept.~00-24 & $1.25\pm 0.14$ & BPL: $\Gamma_1=1.8\pm 0.2$ & {} & $446$ & Day of the discovery \\
{}                 &                & $\Gamma_2=3.2\pm 1.2$      & {}             &     {} & {}\\
{}                 &                & $E_{\rm b}=2.3\pm 1.1$~GeV      & {}             &     {} & {}\\
$4$~Sept.~06-12 & $1.74\pm 0.36$ & PL: $\Gamma=2.0\pm 0.2$ & $3.48\pm 0.72$ & $141$ & Outburst peak flux\\
$1-31$~August & $0.045\pm 0.009$ & PL: $\Gamma=2.1\pm 0.1$ & $0.98\pm 0.09$ & $92$ & Quiescence\\
$1-31$~August & $0.050\pm 0.010$ & BPL: $\Gamma_1=2.2\pm 0.3$ & {} & $87$ & Quiescence \\
{}                 &                & $\Gamma_2=1.9\pm 0.5$      & {}             &     {} & {}\\
{}                 &                & $E_{\rm b}=1.6\pm 1.5$~GeV      & {}             &     {} & {}\\
\tableline
\tableline
\end{tabular}
\end{center}
(a) Observation start-stop [UTC].\\
(b) Flux in the $0.2-300$~GeV energy band [$10^{-6}$~ph~cm$^{-2}$~s$^{-1}$].\\
(c) Spectral shape. PL: power-law model; $\Gamma$ indicates the photon index given by $F(E) \propto E^{-\Gamma}$. BPL: broken power-law model, $\Gamma_1$ is photon index for energies below the break energy $E_{\rm b}$ and $\Gamma_2$ is the photon index for energies greater than $E_{\rm b}$.\\
(d) Flux in the $0.1-300$~GeV energy band extrapolated from the $0.2-300$~GeV flux [$10^{-6}$~ph~cm$^{-2}$~s$^{-1}$].\\
\normalsize
\end{table*}

The LAT data of PKS~1454-354 were fitted with a power-law model for the spectrum (Fig.~\ref{spectra}). On $4$~September~$2008$, the blazar was detected at a flux $(1.29\pm 0.15)\times 10^{-6}$~ph~cm$^{-2}$~s$^{-1}$ in the $0.2-300$~GeV energy band, with a peak of $(1.7\pm 0.4)\times 10^{-6}$~ph~cm$^{-2}$~s$^{-1}$ between $06$ and $12$~UTC. Subsequently, the source was observed in a slower declining phase reaching a flux $(0.56\pm 0.24)\times 10^{-6}$~ph~cm$^{-2}$~s$^{-1}$ after about $18-24$ hours. In Fig.~\ref{lcurves} (\emph{upper panel}) we present the corresponding gamma-ray light curve integrated over $6$-hour time bins between $3$~September~$00$~UT to $6$~September~$2008$~$00$~UT (MJD 54712 - 54715). 

The photon index $\Gamma$, defined as $F(E) \propto E^{-\Gamma}$, averaged over the day, is $2.0\pm 0.1$ and it remains constant through the flare (see Fig.~\ref{lcurves}, \emph{lower panel}). 

Fitting the data with a broken power-law model, does not improve the fit quality implying that no break is detected in the energy range $0.2-5$~GeV (see Fig.~\ref{spectra}). The results of the LAT analysis are summarized in Table~\ref{tab:summary}. The quoted errors are just statistical ones. The systematic errors should be added. According to the studies on the Vela Pulsar (Abdo et al. 2008), our current conservative estimates of systematic errors are $<30$\% for flux measurements and $0.1$ for the photon index. Significant reduction of such systematic uncertainties is expected once the calibration of the LAT instrument will be completed. 

The results of the LAT spectral analysis were confirmed with two additional independent methods: the first one uses a spectrum unfolding based on the Bayes' theorem (D'Agostini 1995), while the second one uses the nested HEALpix (G\'orski et al. 2005). 

\subsection{Swift}
PKS~1454-354 was observed by \emph{Swift} on $7$~September~$2008$ at 11:50~UTC (ObsID~$00036799003$, exposure $1.5$~ks). Another observation performed on $1$~January~$2008$ at 00:16~UTC (ObsID~$00036799001$, exposure $9.0$~ks) is available in the archive. The data from the hard X-ray detector BAT (Barthelmy et al. 2005), the X-ray telescope XRT (Burrows et al. 2005) and the optical/ultraviolet monitor UVOT (Roming et al. 2005) were processed and analyzed with \texttt{HEASOFT v 6.5} and the calibration database updated on $31$~July~$2008$. The results are summarized in Table~\ref{tab:swift}.

The source is not detected by the BAT instrument, even after summing the two observations (exposure $11$~ks). The upper limit ($3\sigma$) is $6\times 10^{-10}$~erg~cm$^{-2}$~s$^{-1}$ in the 20-100~keV band. This is not surprising given the low exposure.

XRT operated in photon-counting mode. Its data were processed with the \texttt{xrtpipeline} task using standard parameters and selecting single to quadruple pixel events (grades 0-12). The output spectra were rebinned to have at least $20$ counts per energy bin. The X-ray spectrum measured on $7$~September is best fitted ($\chi^2=1.3$, dof$=1$) with a power law model with a photon index $\Gamma=2.0\pm 0.5$ with Galactic absorption ($N_{\rm H}=6.46\times 10^{20}$~cm$^{-2}$, Kalberla et al. 2005). We note the very low statistics, which do not allow to put tight constraints, particularly in the photon index. The flux in the 0.2-10~keV energy band was $(2.2\pm 0.2)\times 10^{-12}$~erg~cm$^{-2}$~s$^{-1}$, significantly higher than the value of $(8.9\pm 0.9)\times 10^{-13}$~erg~cm$^{-2}$~s$^{-1}$ measured during the observation performed on $1$~January, indicating that the blazar was still active three days after the $\gamma-$ray outburst detected by LAT. The photon indices of the two observations are, however, consistent within the measurement errors ($\Gamma=2.0\pm 0.3$ on $1$~January). 

UVOT snapshots with all the six available filters, $v$ ($5468$~\AA), $b$ ($4392$~\AA), $u$ ($3465$~\AA), $uvw1$ ($2600$~\AA), $uvm2$ ($2246$~\AA), $uvw2$ ($1928$~\AA), taken during the obsID $00036799003$ observation were integrated with the \texttt{uvotimsum} task and then analyzed by using the \texttt{uvotsource} task, with a source region radius of $5''$ for the optical filters and $10''$ for the UV, while the background was extracted from an annular region centered on the blazar, with external radius of $25''$ and internal radius of $7''$ for the optical filters and $12''$ for the UV. It was not possible to select a larger external radius because of nearby sources. The observed magnitudes were $m_{v}=16.79\pm 0.09$, $m_{b}=17.58\pm 0.06$, $m_{u}=16.79\pm 0.05$, $m_{uvw1}=17.11\pm 0.08$, $m_{uvm2}=17.3\pm 0.2$, $m_{uvw2}=18.0\pm 0.1$. These values were dereddened according to the extinction laws of Cardelli et al. (1989) with $A_{V}=0.34$ and then converted into flux densities according to the standard formulae and zeropoints (Poole et al. 2008). During the observation of $1$~January only one filter ($uvm2$) was available and the measured magnitude was $18.37\pm 0.05$, about one magnitude greater than the value measured on $7$~September.

\begin{table*}
\begin{center}
\scriptsize
\caption{Summary of the \emph{Swift} data. See the text for details.\label{tab:swift}}
\begin{tabular}{cccccccc}
\tableline\tableline
\multicolumn{7}{c}{BAT}\\
\tableline
\multicolumn{3}{c}{ObsID} & Exposure & \multicolumn{3}{c}{Flux$_{20-100 \rm keV}$}\\
\multicolumn{3}{c}{}      & [ks]     & \multicolumn{3}{c}{[$10^{-10}$~erg~cm$^{-2}$~s$^{-1}$]}\\
\tableline
\multicolumn{3}{c}{$00036799001/3$ (sum)} & $11$ & \multicolumn{3}{c}{$< 6.0 $} \\ 
\tableline
\multicolumn{7}{c}{XRT}\\
\tableline
\multicolumn{2}{c}{ObsID} & Exposure & $N_{\rm H}$ & $\Gamma$ & Flux$_{0.2-10 \rm keV}$ & $\chi^{2}$/dof\\
\multicolumn{2}{c}{}    & [ks]     & [$10^{20}$~cm$^{-2}$] & {} & [$10^{-12}$~erg~cm$^{-2}$~s$^{-1}$] & {}\\
\tableline
\multicolumn{2}{c}{$00036799001$} & $9.0$ & $6.46$ & $2.0\pm 0.3$ &  $0.89\pm 0.09$	& $8.0/6$\\
\multicolumn{2}{c}{$00036799003$} & $1.5$ & $6.46$ & $2.0\pm 0.5$ &  $2.2\pm 0.2$   & $1.3/1$\\
\tableline
\multicolumn{7}{c}{UVOT}\\
\tableline
ObsID & $v$ & $b$ & $u$ & $uvw1$ & $uvm2$ & $uvw2$\\
{} & [$5468$~\AA] & [$4392$~\AA] & [$3465$~\AA] & [$2600$~\AA] & [$2246$~\AA] & [$1928$~\AA] \\
\tableline
$00036799001$ & {} & {} & {} & {} & $18.37\pm 0.05$ & {} \\
$00036799001$ & $16.79\pm 0.09$ & $17.58\pm 0.06$ & $16.79\pm 0.05$ & $17.11\pm 0.08$ & $17.3\pm 0.2$ & $18.0\pm 0.1$\\
\tableline
\tableline
\end{tabular}
\normalsize
\end{center}
\end{table*}

\subsection{Ground-based Optical Observations}
Optical observations in Johnson $R$ and $B$ filters were obtained on $6$~September at 17:44~UTC and $7$~September at 18:30~UTC with the $0.8$~m optical telescope \emph{ATOM} in Namibia. \emph{ATOM} is operated robotically and obtains automatic observations of confirmed or potential gamma-bright blazars. Data analysis (de-biassing, flatfielding, photometry using \texttt{SExtractor}, Bertin \& Arnouts, 1996) is conducted automatically. PKS~1454-354 does not appear to be spatially extended beyond the point-spread function and no corrections for potential contributions from the host galaxy were performed. Absolute photometry was obtained using non-variable field stars in photometric nights using several calibration stars. The observed magnitudes are: $B=17.01\pm 0.02$ and $R=16.46\pm 0.02$ on $6$~September, and $B=16.83\pm 0.02$ and $R=16.53\pm 0.02$ on $7$~September. The magnitudes were then corrected for galactic extinction using the same methods outlined above for \emph{Swift}/UVOT and converted in fluxes using standard formulae and zeropoints. Comparison with \emph{Swift}/UVOT data indicates some intraday variability.

\begin{figure*}
\centering
\includegraphics[scale=0.5]{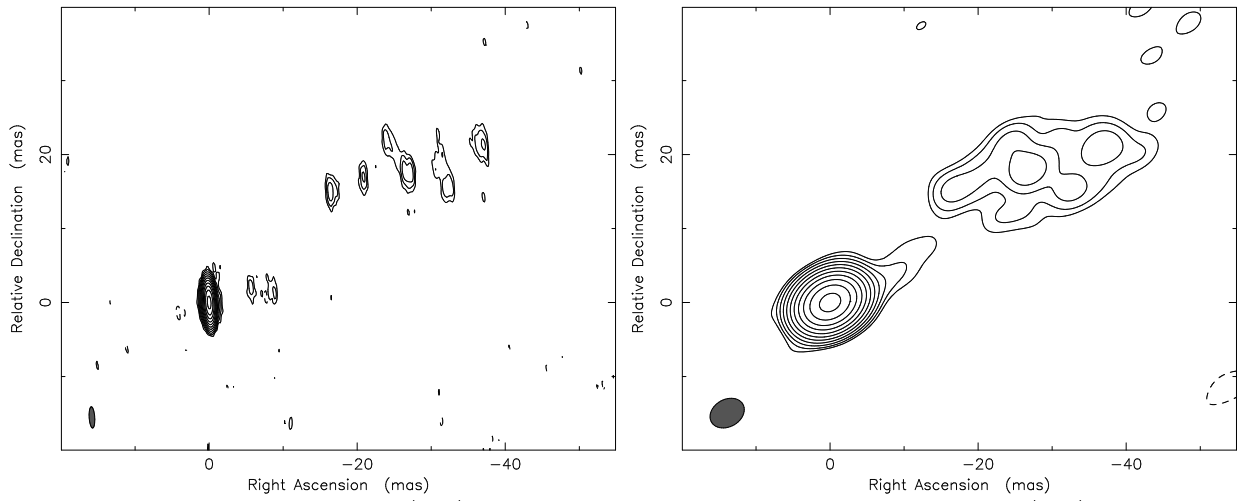}
\caption{VLBI images of PKS~1454-354 at $8.4$~GHz. The left panel shows the naturally weighted image and the right panel the tapered image obtained from imaging the final calibrated data with down-weighted visibilities of long baselines ($10$\% at $100$ million wavelengths) to increase the signal-to-noise ratio of the low-surface-brightness jet. The jet extends to $\sim 50$~mas to the west-northwest. Both images have a rms noise of $\sim 0.08$~mJy/beam and the lowest contour is at $3.7$ and $3.8$ times the rms noise, respectively. The restoring beam is shown on bottom left of each image ($2.87 \times 0.75$~mas at $4^{\circ}$; $4.9 \times 3.7$~mas at $-60^{\circ}$). These data were obtained on $10$~November~$2007$, as part of the TANAMI program (Tracking AGN with Austral Milliarcsecond Interferometry; Ojha et al. 2008) with six elements of the Australian South-African LBA (Long Baseline Array). See the text for details.}
\label{radio}
\end{figure*}

\subsection{Radio Observations}
Shortly after detecting the high $\gamma-$ray flux reported in this paper, on $13$~September~$2008$, PKS~1454-354 was observed with the \emph{Australia Telescope Compact Array} (ATCA) at $2.368$, $4.800$, $8.640$, $18.496$ and $19.520$~GHz. At this time the array was in its H75 configuration, with a maximum separation between the inner five antennas of $90$~m and the sixth antenna located $4.3$~km to the west. The inner five antennas were located on the main east-west track and north-south spur, providing instantaneous two-dimensional $(u,v)$ coverage.

The ATCA primary calibrator, PKS~1934-638, was observed first and used to calibrate the array at all frequencies over the $128$~MHz observing bandwidth. The data were reduced following the standard procedures in MIRIAD, and the task \texttt{uvflux} used to fit a point source model to the data and determine the flux density and rms scatter about this point. Inspection of the data confirmed that the point source approximation was valid at all frequencies. Phase stability was poor on the long baselines to the sixth antenna, particularly at the higher frequencies and so this antenna was omitted from the analysis. The fitted flux densities were $0.76\pm 0.08$, $0.77\pm 0.02$, $0.87\pm0.02$, $1.23\pm0.07$, $1.28\pm0.08$~Jy at $2.368$, $4.800$, $8.640$, $18.496$ and $19.520$~GHz, respectively. The quoted errors are statistical: in addition there is a systematic error of about $2$\% (see, e.g., Tingay et al. 2003).

These values can be compared with cataloged flux densities for this source. PKS~1454-354 was included in the AT20G Bright Source Survey (Massardi et al. 2008) with a flux density at $19.904$~GHz of $0.90\pm0.05$~Jy determined from an observation in May~$2007$, indicating the source had brightened by $40$\% by September~$2008$. The PMN survey (Wright et al. 1996) recorded a flux density of $0.566\pm0.041$~Jy at $4.85$~GHz in November~$1990$. It can be concluded that, at the time of the $\gamma-$ray outburst, PKS~1454-354 was in a comparatively bright radio state showing an inverted spectrum, which can be interpreted as the indication of an enhanced ongoing radio activity. Single dish radio flux-density monitoring observations are ongoing at the Ceduna $30$~m telescope and will be reported elsewhere.

Fig.~\ref{radio} shows VLBI images of PKS~1454-354 at a frequency of $8.4$~GHz. These data were obtained on $10$~November~$2007$, as part of the TANAMI program (Tracking AGN with Austral Millarcsecond Interferometry; Ojha et al. 2008) with the Australian Long Baseline Array consisting of telescopes at Parkes, Narrabri, Mopra, Hobart, Ceduna (all in Australia) and Hartebeesthoek (South Africa). The data were correlated using the \texttt{DiFX} software correlator (Deller et al. 2007) at the Swinburne University of Technology. Details of these telescopes as well as the data calibration path are described in Ojha et al. (2005). The approximate position angle of the diffuse jet is $\sim 300^\circ$. The correlated flux density of the milliarcsecond-scale core-jet structure is about a factor of 2 below the (non-simultaneous) total flux density at $8.64$~GHz measured by ATCA in September~$2008$: $\sim 415$~mJy, of which more than $95$\% is contained within the compact core, i.e., on scales smaller than $0.75$~mas (the full width at half maximum of the minor axis of the beam). Model fitting the core with an elliptical Gaussian function results in a model component with $400$~mJy and a size of $0.26\times 0.17$~mas, corresponding to a brightness temperature limit of $T_{\rm B} = 1.6 \times 10^{11}$~K. These are the first published VLBI images of PKS~1454-354 at this angular resolution and dynamic range. The only other radio structural information on these scales comes from the VCS3 program (Petrov et al. 2005) carried out in May~$2004$ and showing no structure at $8.4$~GHz and a jet to the west-northwest at $2.3$~GHz consistent with the morphology that we find at $8.4$~GHz. Petrov et al. (2005) report $0.51$~Jy at both $2.3$ and $8.4$~GHz, and unresolved components of $0.38$ and $0.21$~Jy respectively.

\section{Discussion of the SED}
\begin{figure}
\centering
\includegraphics[angle=270,scale=0.35]{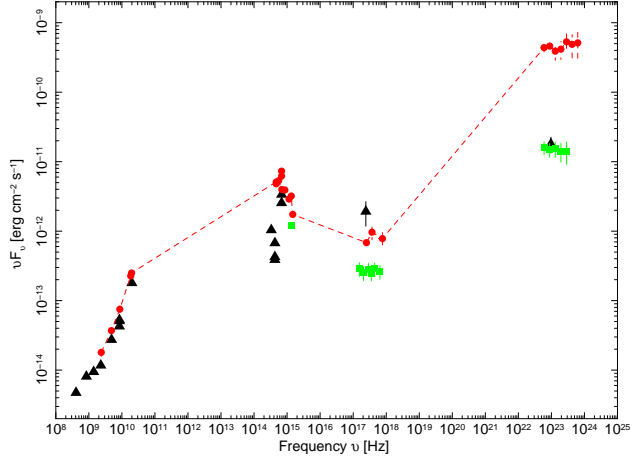}
\caption{Spectral Energy Distribution of PKS~1454-354. \emph{ATCA} ($13$~September), \emph{ATOM} ($6$ and $7$~September), \emph{Swift} (XRT and UVOT; $7$~September) and \emph{Fermi} (LAT; $4$~September) data are indicated with filled circles (red) and connected with a dashed line for a better visualization. \emph{Fermi}/LAT data of August and \emph{Swift} (XRT and UVOT) observation of $1$~January~$2008$, to which we refer as ``low state'', are indicated with squares (green). Archival data are marked with filled triangles (black). Radio data: Molonglo, $408$~MHz (Large et al. 1981); NVSS, $1.4$~GHz (Condon et al. 1998); Parkes, $4.85$~GHz (Wright et al. 1996);  CRATES, $8.4$~GHz (Healey et al. 2007); ATCA, $20$~GHz (Massardi et al. 2008); SUMSS, $843$~MHz (Mauch et al. 2003); VCS3, $2.3$ and $8.6$~GHz (Petrov et al. 2005). Optical: USNO B1, $B$, $R$, $I$ filters (Monet et al. 2003); Guide Star Catalog II, $B$ and $R$ filters (Lasker et al. 2008). X-rays: ROSAT Faint Source Catalog, $1$~keV  (Voges et al. 2000). Gamma-Rays: EGRET, $400$~MeV (Hartman et al. 1999).}
\label{sed}
\end{figure}

In Fig.~\ref{sed} we report the spectral energy distribution (SED) built with archival data together with \emph{Fermi}/LAT, \emph{Swift}, \emph{ATCA} and \emph{ATOM} data analyzed in the present work. Two different states are shown for $\gamma-$rays (LAT): the quiescence value, measured from the 1-month data set, and the active state, obtained from the data collected on $4$~September, the day of the discovery. A change in flux during the flare of more than one order of magnitude is evident.

Fluxes at X-rays, ultraviolet, optical and radio frequencies are higher than those of the archival data, although the \emph{Swift} observation was performed on $7$~September and \emph{ATCA} measurements were done on $13$ September. This confirms that PKS~1454-354 was still active at all the wavelengths several days after the $\gamma-$ray flare observed by LAT. It is worth noting the absence of significant spectral changes in the X- and $\gamma-$ray bands, but given the low statistics, we cannot exclude the presence of spectral variations on short time scales. However, at radio frequencies, there was an inversion of the spectral index $\alpha$, defined as $S_{\nu}\propto \nu^{-\alpha}$. The value calculated from archival data is positive ($\alpha_{1.4\rm GHz}^{8.4\rm GHz}=0.040$, Sowards-Emmerd et al. 2004; $\alpha_{1.4\rm GHz}^{8.6\rm GHz}=0.054$, Healey et al. 2007; with an error of about $7$\% in both cases) and becomes negative during the outburst ($\alpha_{2.368\rm GHz}^{8.640\rm GHz}=-0.10\pm 0.01$). This can be interpreted as the signature of the emission of a relativistic plasma blob.

In order to investigate the SED characteristics and compare with larger samples of similar objects, 2-point spectral indices have been calculated according to the formula of Ledden \& O'Dell (1985). We used radio data at $4.8$~GHz, optical data at $\lambda=3465$~\AA (\emph{Swift}/UVOT $u$ filter, to minimize the contribution of the host galaxy) and X-rays at $1$~keV. From ATCA radio measurements of $13$~September, optical/UV observations from \emph{Swift}/UVOT and X-rays from \emph{Swift}/XRT collected on $7$~September, we calculated $\alpha_r=-0.1$, $\alpha_o=1.85$ and $\alpha_x=1.0$, respectively. Then, the flux densities have been K-corrected by the multiplicative factor $(1+z)^{\alpha-1}$. Finally, we obtain the following values: $\alpha_{ox}=1.11$, $\alpha_{ro}=0.49$, $\alpha_{rx}=0.69$. 

These values, particularly $\alpha_{ro}<0.5$, seem to suggest that this FSRQ belongs to the class of high-frequency peaked FSRQ (HFSRQ) conjectured by Padovani et al. (2003), which have the synchrotron emission peaked at UV/soft X-rays. However, the shape of the SED, specifically the strong change in spectral index between optical and X-ray frequencies ($\alpha_o=1.85$ and $\alpha_x=1.0$), clearly rules out this possibility.

Instead, our 2-point spectral indices are very similar to those of X-ray selected FSRQ analyzed in Maraschi et al. (2008), where, in addition to the usual non-thermal model of blazars (synchrotron self-Compton, SSC, plus external Compton, EC), there is a component due to the X-ray corona above the accretion disk (for the disk-jet connection in FSRQ see also Grandi \& Palumbo 2004, Sambruna et al. 2007). The moderately high value of $\alpha_{ox}$, together with a soft X-ray spectrum ($\alpha_{x}=1$), suggests the presence of a corona, which is best observed when the source is in quiescence and the jet contribution is low. The contrary occurs during $\gamma-$ray flares, due to the dominance of the jet emission over the corona. In this case, the X-ray spectrum should become harder ($\alpha_{x}<1$), but the measured X-ray photon index has a large error ($\Gamma = 2.0 \pm 0.5$) and, although consistent with this hypothesis, it is not conclusive.

To support this hypothesis, we calculate some order-of-magnitude values for the physical parameters and we refer to Maraschi et al. (2008) for more details. These authors found that $\alpha_{ox}$ is correlated with the viewing angle and, therefore, is anticorrelated with the Doppler factor ($\delta$). Taking advantage of such correlation, we can estimate a value of $\delta=15-20$, which in turn allows us to calculate the size of the emitting region in the source frame $R'$. According to the well-known formula $R'<c\Delta t\delta/(1+z)$, the emitting region must be smaller than $R'\lesssim 10^{16}$~cm during the rise of the $\gamma-$ray flare, whose timescale is estimated to be $\approx 12$~hours (see Fig.~\ref{lcurves}). The longer timescale during the declining part of the flare suggests a larger emitting region, meaning that the electrons escaped from the processing region, resulting in a decrease of the jet power. A similar effect can be obtained decreasing the density of seed photons, but in this case we expected a change in the optical emission too, disentangled from the synchrotron emission. However, the coordinated increase of the whole synchrotron hump, from radio to optical frequencies, makes the change in the injected power the most likely explanation for the observed variability. 

The jet-frame radiation energy density can be estimated by the energy density of the broad-line region (BLR) according to the formula $U_{\rm rad}\sim U_{\rm BLR}\delta^2$ (given that the Doppler factor is almost equal to the bulk Lorentz factor). Since the BLR luminosity is assumed to be about $10\%$ of $L_{\rm disk}$ ($\sim 4\times 10^{46}$~erg~s$^{-1}$ in our case), the size of BLR can be calculated as $R_{\rm BLR}\sim 10^{17}\sqrt{L_{\rm disk,45}}=6\times 10^{17}$~cm, where $L_{\rm disk,45}$ is the disk luminosity in units of $10^{45}$~erg~s$^{-1}$. Then, the BLR energy density is $U_{\rm BLR}=L_{BLR}/(4\pi c R^2_{\rm BLR})=0.029$~erg~cm$^{-3}$. The corresponding total radiation energy density in the jet-frame is therefore $U_{\rm rad}\sim 6-12$~erg~cm$^{-3}$, depending on the value of $\delta$, which, in turn, results in a peak electron Lorentz factor of the order of $\gamma_{\rm peak}\sim 100-1000$. These values are close to the parameters obtained with a more detailed modeling of similar FSRQ reported in Maraschi et al. (2008). 

The large errors, particularly in the X-ray photon index, leave open different possibilities, involving one or two SSC components (e.g. Ballo et al. 2002, Giommi et al. 2008). In addition, flat X-ray spectra from blazar jets may possibly be a signature of pair cascades (e.g., Svensson 1987, M\"{u}cke et al. 2003, B\"{o}ttcher et al. 2008).

\section{Final Remarks}
We have reported the discovery by \emph{Fermi}/LAT of $\gamma-$ray emission from the FSRQ PKS~1454-354. LAT data showed flux variability on $6$ to $12$ hours timescale, but with negligible spectral changes. The blazar is positionally consistent with the unidentified source 3EG~J1500-3509 and, therefore, it is likely to be the EGRET source counterpart. The increased activity measured at radio, optical, ultraviolet and X-ray frequencies in the days following the $\gamma-$ray outburst strengthens the identification of PKS~1454-354 as a high-energy $\gamma-$ray source.

The analysis of the SED and some order-of-magnitude calculations suggest that PKS~1454-354 is a typical FSRQ, although the large measurement errors hampered our capability to put tight constraints on the origin of the X-ray emission: the soft X-ray spectrum can be explained either in terms of SSC or of a corona above the accretion disk, while the high-energy peak is likely to be due to external Compton processes. However, other theories (e.g., hadronic models) can also be considered to explain the observed data. 

Further, more precise multiwavelength observations of PKS~1454-354 and, specifically, a spectral coverage at sub-mm, IR and optical frequencies, where the synchrotron peak is expected, will shed further light on the modeling of this source, allowing a more precise correlation between the synchrotron and $\gamma$-ray inverse-Compton peak in quiet and flaring states, clarifying the astrophysical mechanisms at work in this source.

\acknowledgments
The \emph{Fermi}/LAT Collaboration acknowledges generous ongoing support from a number of agencies and institutes that have supported both the development and the operation of the LAT as well as scientific data analysis.  These include the National Aeronautics and Space Administration and the Department of Energy in the United States, the Commissariat \`a l'Energie Atomique and the Centre National de la Recherche Scientifique / Institut National de Physique Nucl\'eaire et de Physique des Particules in France, the Agenzia Spaziale Italiana and the Istituto Nazionale di Fisica Nucleare in Italy, the Ministry of Education, Culture, Sports, Science and Technology (MEXT), High Energy Accelerator Research Organization (KEK) and Japan Aerospace Exploration Agency (JAXA) in Japan, and the K.~A.~Wallenberg Foundation, the Swedish Research Council and the Swedish National Space Board in Sweden.

Additional support for science analysis during the operations phase from the following agencies is also gratefully acknowledged: the Istituto Nazionale di Astrofisica in Italy and the K.~A.~Wallenberg Foundation in Sweden for providing a grant in support of a Royal Swedish Academy of Sciences Research fellowship for JC.

The Australia Telescope Compact Array and Long Baseline Array are part of the Australia Telescope which is funded by the Commonwealth of Australia for operation as a National Facility managed by CSIRO. We would like to thank the rest of the TANAMI team for their efforts that led to the observations reported here.

This research has made use of the NASA/IPAC Extragalactic Database (NED) which is operated by the Jet Propulsion Laboratory, California Institute of Technology, under contract with the National Aeronautics and Space Administration. 
This research has made use of data obtained from the High Energy Astrophysics Science Archive Research Center (HEASARC), provided by NASA's Goddard Space Flight Center. \\

\end{document}